# An Optical Lattice Clock with Accuracy and Stability at the $10^{-18}$ Level

B. J. Bloom[1,2,*], T. L. Nicholson[1,2,*], J. R. Williams[1,2,♣], S. L. Campbell[1,2], M. Bishof[1,2], X. Zhang[1,2], W. Zhang[1,2], S. L. Bromley[1,2], and J. Ye[1,2]

**Abstract**

The exquisite control exhibited over quantum states of individual particles has revolutionized the field of precision measurement, as exemplified by the most accurate atomic clock realized in single trapped ions. Whereas many-atom lattice clocks have shown advantages in measurement precision over trapped-ion clocks, their accuracy has remained 20 times worse. Here we demonstrate, for the first time, that a many-atom system achieves accuracy ($6 \times 10^{-18}$) better than a single ion-based clock, with vastly reduced averaging times (3000 s). This is the first time a single clock has achieved the best performance in all three key ingredients necessary for consideration as a primary standard - stability, reproducibility, and accuracy. This work paves the way for future experiments to integrate many-body quantum state engineering into the frontiers of quantum metrology, creating exciting opportunities to advance precision beyond the standard quantum limit. Improved frequency standards will have impact to a wide range of fields from the realization of the SI units, the development of quantum sensors, to precision tests of the fundamental laws of nature.

[1]JILA, National Institute of Standards and Technology and University of Colorado, Boulder, CO 80309-0440, USA.
[2]Department of Physics, University of Colorado, Boulder, CO 80309-0390, USA.
[*]These authors contributed equally towards this work
[♣]Present address: Jet Propulsion Laboratory, California Institute of Technology, Pasadena, CA, USA.

**Introduction**

The rapid improvement in atomic clocks has benefited from outstanding progress in atomic, optical, and quantum science[1,2]. At the same time, research on the best atomic clocks has contributed strongly to advancing the frontiers of science, with strong impacts to both fundamental and applied research. The realization and distribution of the SI units [3], search for time variation of fundamental constants [4], high-resolution XUV spectroscopy [5], and relativistic geodesy will all benefit from the expanded use of next generation frequency standards [6]. Furthermore, new advances in frequency and time base distribution lend credence to the extended utility of advanced frequency standards outside the lab [7,8].

Prior to this work, owing to the exquisite control of quantum states of trapped ions, optical clocks based on single particles held the record for the lowest systematic uncertainty [♦,4,9–12]. However, neutral atom clocks with many ultracold atoms confined in magic-wavelength optical lattices [13] have the potential for much greater precision than ion clocks [14–17]. This potential has been realized only very recently owing to the improved frequency stability of optical local oscillators [18–20], resulting in a record single-clock instability of $3.1 \times 10^{-16} /\sqrt{\tau}$ [21], where $\tau$ is the averaging time in seconds. This result represents a gain by a factor of 10 in our clock stability, allowing for a factor of 100 reduction in the averaging time needed to reach a desired uncertainty [21]. Equivalent instability has also been achieved with two Yb optical lattice clocks, but it was demonstrated over a longer averaging time of 7 hours [22]. We have used this measurement precision to systematically evaluate important effects that have limited optical lattice clocks, achieving a total systematic uncertainty in fractional frequency of $6.4 \times 10^{-18}$, which

---

♦ Accuracy for the SI second is currently defined by the Cs primary standard. A number of optical atomic clocks have achieved lower systematic uncertainty than that of the Cs clock. This systematic uncertainty becomes accuracy once a redefinition of the SI second is made based on an optical atomic clock.

is a factor of 20 improvement over the best reported total uncertainties for optical lattice clocks [15,17,23,24] and is now the best among all atomic clocks.

Such systematic evaluations for critical parameters can be routinely performed in 3000 s, owing to the clock's instability of $3\times10^{-18}$ at 10,000 s. Combining such precision with improved clock designs has allowed us to overcome two main obstacles and achieve the breakthroughs in uncertainty reported here. First, we must understand and overcome atomic-interaction-induced frequency shifts inherent to many-particle clocks [25–27]. We have now determined this effect with $6\times10^{-19}$ uncertainty. Second, we must accurately measure the thermal radiation environment of the lattice-trapped atoms as this causes the largest systematic clock shift, known as the blackbody radiation (BBR) Stark shift. Incomplete knowledge of the thermal radiation impinging upon the atoms has so far limited lattice clock uncertainty. We demonstrate that a combination of accurate *in-situ* temperature probes and a thermal enclosure surrounding the clock vacuum chamber has allowed us to achieve an overall BBR shift uncertainty of $4.1\times10^{-18}$. This progress was enabled by a precise measurement (performed at PTB) of the Sr polarizability [28], which governs the magnitude of the BBR shift. Furthermore, two independent Sr clocks are compared and they agree within their combined total uncertainty over a period of one month. These results herald a new age in frequency metrology where optical lattice clocks exhibit the best performance in both stability and total uncertainty.

**Frequency Comparison**

To demonstrate the improved performance of lattice clocks, we have built two Sr clocks in JILA [21,29]. Here we refer to the first generation JILA Sr clock as SrI and the newly constructed Sr clock as SrII. The recent improvement of low thermal noise optical oscillators has allowed us to demonstrate the stability of both Sr clocks reaching within a factor of 2 of the quantum projection noise (QPN) limit for 2000 atoms [21]. We constructed the SrII clock with the goal of reducing the atomic interaction- and BBR-related frequency uncertainties. Thus, SrII has an optical trap volume ~100 times larger than SrI to reduce the atomic density, along with *in-situ* BBR probes in vacuum to measure the thermal environment of the atoms. The improvement of SrI, on the other hand, has been a modest factor of 2 over our previous result [15], now achieving a total systematic uncertainty of $5.3 \times 10^{-17}$. When the SrI and SrII clocks were compared over a period of one month, we found their frequency difference to be $\nu_{SrII} - \nu_{SrI} = -2.8(2) \times 10^{-17}$, well within their combined uncertainty of $5.4 \times 10^{-17}$.

For both clocks, up to a few thousand $^{87}$Sr atoms are laser cooled to about 1 μK and loaded into one-dimensional optical lattices near the magic wavelength. A laser with a short-term stability of $1 \times 10^{-16}$ (from 1 to 1000 s) interrogates the $^1S_0 - {}^3P_0$ clock transition with a coherent spectroscopy time of 160 ms. We normally operate the clock comparison in an asynchronous interrogation mode, where the two clock probe pulses are purposely non-overlapping in time [21]. Two independent frequency shifters are used to correct the laser frequency to the SrI and SrII clock transitions, respectively. State detection of both atomic ensembles is a destructive measurement that requires the repetition of the experimental cycle every 1.3 s. After each cycle, the frequency corrections, atom numbers, and environmental temperatures for both systems are recorded and time stamped for comparison and post-processing.

A major practical concern is the speed with which these clocks reach agreement at their stated uncertainties. Hence, the stability of these Sr clocks, displayed as the Allan deviation of their frequency comparison in Fig. 1a, is critical for evaluating systematic effects in a robust manner. Figure 1b documents the reproducibility of both SrI and SrII clocks, where a comparison of their frequencies over a period of one month shows that they disagree at the level of $2.8(2)\times10^{-17}$. The Allan deviation and the binned intercomparison data showcase the stability and reproducibility of these clocks on the short- and long-term timescales. This performance level is necessary for a rigorous evaluation of clock systematics at the $10^{-18}$ level.

SrI and SrII independently correct for systematic offsets to their measured atomic frequencies. Table 1 lists the major sources of frequency shifts (in columns labeled with $\Delta$) and their related uncertainties (in columns labeled with $\sigma$) that affect both clocks. The SrI clock uncertainty is dominated by its BBR shift uncertainty of $4.5\times10^{-17}$. For SrII, on the other hand, all sources of uncertainties have been evaluated better than $4\times10^{-18}$. Below we provide a detailed discussion on some of the main systematic effects.

**Blackbody Radiation**

Compared to other lattice clocks, the largest improvement in the total systematic uncertainty of SrII is the control of the BBR shift. We enclose the entire clock apparatus inside a black box (Fig. 2a). Our lasers for cooling, trapping, and clock spectroscopy are delivered to the inside of the black box by optical fibers, preventing stray radiation from entering. We have also installed *in-situ* silicon diode temperature sensors near the optical lattice in the vacuum to measure the radiative heat environment seen by our atoms. Silicon diode sensors are utilized for their ease of calibration (as their forward voltage drop is linear in temperature) and their suitability for vacuum baking [30]. Two such sensors (with NIST-traceable calibrations) are affixed to separate glass tubes (Fig. 2a). The glass tubes prevent parasitic heat conduction from the chamber to the sensors by providing insulation and radiative dissipation of conductive heat. To improve the radiative coupling, surfaces of the sensors are coated with high absorptivity, UHV-compatible paint. One sensor is mounted 1 inch away from the atomic cloud and provides real time temperature monitoring during clock operation. The second sensor is affixed to an in-vacuum translator, allowing for mapping of the temperature gradients near the lattice confined atoms (Inset, Fig. 2c). During clock operations the mechanical translator can be retracted to avoid interference with the atomic cloud, as can be seen in Fig. 2b. Systematic errors in both the readout of the sensors and their ability to determine the actual thermal distribution at the position of the atoms result in an overall uncertainty of 26.7 mK for the stated BBR temperature. Table 2 lists the sources of uncertainties for this temperature evaluation.

The atoms, unlike the temperature sensors, are influenced not only by the total integrated power of the BBR inside the chamber, known as the BBR static correction, but also the frequency-weighted spectrum of the radiation inside the chamber, known as the BBR dynamic correction. We constructed a ray tracing model of our chamber to estimate the influence of temperature gradients throughout the vacuum chamber [31,32]. The model predicts the error incurred in the BBR dynamic correction by calculating the deviation from a perfect BBR spectrum at the temperature read out by the sensor[33]. Due to the large difference in emissivity between the vacuum viewports and the metal chamber, the model is insensitive to the absolute emissivity values of the elements. As seen in Fig. 2c, components that couple strongly to the sensor such as the large viewports would need to deviate in temperature from the rest of the chamber by a significant amount (more than 10 K) for this error to reach the $1\times10^{-18}$ level. However, much hotter components which couple more strongly to the atoms, such as the heated Zeeman slower window, introduce a larger error in the deviation from a perfect BBR spectrum. For the systematic uncertainty evaluation, the operation of the Sr optical lattice clock can be performed without heating the Zeeman slower window for a limited amount of time. Even with the Zeeman slower window heated, its contribution to total uncertainty is below $1.2\times10^{-18}$. For future experiments that require total uncertainty below $1\times10^{-18}$, we can simply add mechanical shutters that obscure the atoms' view of all hot elements in the system. We ensure that the black box enclosing the SrII apparatus is fully sealed from the outside environment, forbidding the atoms to view any highly emissive object with a temperature differing from the ambient temperature inside the box. This BBR shielding box also allows the clock vacuum chamber to be decoupled from room temperature variations as it reaches an equilibrium temperature of 301 K after two hours of clock operation.

**Atomic Trapping Effects**

All other systematics listed in Table 1 are rapidly measured through self-comparison with digital lock-in [21,34]. Both SrI and SrII measure their systematics by modulating a particular physical parameter every two experimental cycles, with the clock laser serving as a stable reference to measure the related frequency shifts. For example, the atomic density shift is measured to a high precision with this method. The SrII system is designed to reach a density shift uncertainty below $1 \times 10^{-18}$ by using large lattice trapping volumes. To accommodate this, a Fabry-Perot buildup cavity is employed to achieve a sufficiently deep lattice. This trap design increases the number of atoms loaded into the lattice at a decreased atomic density, allowing SrII to measure an already reduced density shift to very high precision. Details of the SrI and SrII optical lattice trap geometries can be found in Ref. 21.

Frequency shifts induced by the optical lattice potential must be understood and controlled at an extremely high level of precision, especially for optical lattices that trap weakly against gravity [35]. A variety of methods are employed to stabilize the lattice scalar, vector, and tensor Stark shifts. An 813 nm continuous-wave Ti:sapphire laser is used to create the lattice light for SrII. The clean spectrum of the solid state laser is advantageous over a semiconductor tapered amplifier (TA)-based lattice, where spontaneous noise pedestals might cause additional frequency shifts [17]. For SrI, we use a TA system, but we refine the output spectrum with a narrow band interference filter and an optical filter cavity. To deal with potential residual shifts due to the TA noise pedestals, we regularly calibrate the lattice Stark shift for SrI. Both clocks stabilize their lattice laser frequencies to a Cs clock via a self-referenced Yb fiber comb, and their trapping light intensities are stabilized after being delivered to the atoms. The lattice vector shift is cancelled by alternately interrogating the +9/2 and -9/2 stretched nuclear spin states of the atom on successive experimental cycles, in addition to the use of linearly polarized lattice light [15,35]. This interrogation sequence also allows cancellation of the first order Zeeman shift. Rather than trying to artificially separate the scalar and tensor shifts, we can treat them as a single effect in our measurement of the AC Stark shift [35]. We further minimize the tensor shift's sensitivity to the magnetic bias field (B) by setting the lattice polarization and the direction of B to be parallel. Modulating the intensity of the lattice, we do not identify any lattice shifts that are nonlinear in lattice intensity. A Fisher test (or F-test) performed for various model shifts on an extensive set of data shown in Fig. 3a demonstrate that the lattice shift is consistent only with a linear model to within our quoted uncertainty.

**Magnetic and Electric DC Fields**

For SrII we also take extra care to minimize fluctuations in the magnetic field-related lattice Stark effect and the second order Zeeman shift [35]. To stabilize the magnetic field for our clock over long operational periods, we use the atoms themselves as a collocated magnetometer for the clock. Every two minutes during the clock operation, the computer-based frequency locking program is paused to interrogate unpolarized atomic samples under a zero applied magnetic field. A drift in the background magnetic field then shows up as a reduced excitation for the peak of an unpolarized line, as all 10 nuclear spin states will experience different Zeeman shifts. Every time the magnetic field servo is activated, the program automatically dithers each pair of magnetic field compensation coils (along three orthogonal spatial directions) and optimizes the current for each pair of coils. As shown in Fig. 3b, the magnetometer-based feedback loop not only keeps the field direction constant throughout the clock operation, but also automatically nulls the field without an operator's intervention. For SrI this procedure is unnecessary due to its more stable magnetic field environment.

To push the systematic evaluation below $1\times10^{-16}$, we also need to evaluate the DC electric-field-induced Stark effect, a frequency shift mechanism caused by patch charges immobilized in the vacuum chamber's fused silica viewports [36]. A pair of disk-shaped electrodes is placed near the two largest viewports in the system (separated along the vertical direction) and shifts are recorded as the electrode polarity is switched. Differences between the frequency shifts induced by oppositely charged electrodes indicate the presence of stray background electric fields as shown in Fig. 3c. When first measuring the DC Stark effect on SrII, a residual $-1.3\times10^{-16}$ shift was discovered. However, when installing the temperature probes, the SrII vacuum chamber was filled with clean nitrogen and then re-evacuated. We found that this process had reduced the measureable DC Stark effect to $-3.5\times10^{-18}$. To complete the full evaluation of the DC Stark effect, we have performed similar measurements along the horizontal direction and found negligible effects below $1\times10^{-18}$.

**Conclusions**

The outlook for the optical lattice clocks is bright. We note that this is only the first systematic evaluation of a lattice clock enabled by the new generation of stable lasers, which led to clock stability near the QPN limit for 1000 atoms. As the laser stability continues to improve

[37], Sr and other lattice clocks will continue to advance by increasing the QPN-limited precision with larger numbers of atoms. Along with the dramatic advances in clock stability, the systematic uncertainty for the clock will rapidly decrease due to the vastly reduced measurement times. Hence future lattice clocks will have their stability and total uncertainty advance in locked steps. The techniques demonstrated in this work have no foreseeable limitations to achieving clock stability and total uncertainty below $1\times10^{-18}$. These advanced clocks will in turn push forward a broad range of quantum sensor technologies and facilitate a variety of fundamental physics tests.

**Acknowledgments:** We thank M. Martin, M. Swallows, E. Arimondo, J. L. Hall, and T. Pfau for useful discussions and H. Green for technical assistance. This research is supported by the National Institute of Standards and Technology, Defense Advanced Research Projects Agency QuASAR Program, and NSF PFC. M.B. acknowledges support from the National Defense Science and Engineering Graduate fellowship program. S.C. acknowledges support from the NSF Graduate Fellowship.

**Author Information:** Any mention of commercial products does not constitute an endorsement by NIST. Correspondence and requests for materials should be addressed to J.Y. (Ye@jila.colorado.edu).

**Fig. 1. Clock comparisons between SrI and SrII. a,** Allan deviation of the SrI and SrII comparison divided by $\sqrt{2}$ to reflect the performance of a single clock. The red solid line is the calculated QPN for this comparison. The green dashed line is a fit to the data, showing the worst case scenario for the averaging of a single clock of $3.4 \times 10^{-16}$ at 1 s. **b,** The absolute agreement between SrI and SrII recorded at the indicated GMT. The light-green region denotes the 1σ combined systematic uncertainty for the two clocks under the running conditions at that time. Each solid circle represents 30 minutes of averaged data. The green dashed lines represent the 1σ standard error inflated by the square root of the reduced $\chi^2$ for the weighted mean of these binned comparison data. The final comparison showed agreement at $-2.8(2) \times 10^{-17}$ with 52,000 seconds of data.

**Fig. 2. Characterizing Black Body Radiation effects on the $^1S_0$-$^3P_0$ transition. a,** A three-dimensional model of the clock vacuum chamber. The sensor mounted on an in-vacuum translator is depicted in its fully extended mode of operation. The entire clock chamber resides inside a thermally isolated enclosure. After two hours of operation, the internal temperature inside the thermally insulating black box becomes decoupled from the room temperature variations and reaches an equilibrium temperature of 301 K. **b,** A photograph of the two glass tubes surrounding a cloud of $^{87}$Sr atoms (red arrow) trapped in a magneto-optical trap formed with 461nm light addressing the $^1S_0$-$^1P_1$ transition. The movable sensor (green arrow) has been retracted to demonstrate its normal running operation. The Conflat pictured is 2.75" in diameter. **c,** The error inherent in assuming a perfect BBR spectrum inside the vacuum chamber, based on a measurement of total BBR radiated power. A perfect BBR environment corresponds to a spatially uniform temperature; deviations from such environment cause a temperature gradient inside the chamber. By modeling all components of the chamber as 301 K and varying the bottom window temperature (shown in the top horizontal axis), one sees that measuring the total radiative power inside the chamber is sufficient for our quoted systematic uncertainty for BBR. The bottom horizontal axis displays the temperature difference between the locations of the atoms and the retracted movable sensor. The inset is a typical measured temperature gradient inside the vacuum chamber referenced to the temperature of the retracted movable sensor, taken by slowly translating the movable sensor from a retracted position (green diamond) to an extended position at the atoms (red square) and back again.

**Fig. 3. Examples of Systematic Evaluations. a,** To accurately determine the correct functional form to ascribe to the AC Stark effect associated with the optical lattice, a variety of lattice trap depths were used to verify its behavior. This effect is depicted as a function of the differential lattice depth. Within our measurement precision, the best fit is a linear model. **b,** Using the atomic cloud as a collocated magnetometer, a residual non-zero magnetic field can be inferred via the peak excitation of an unpolarized Rabi lineshape. The left panel shows the action of our servo as it actuates on 3 pairs of coils to bring the residual magnetic field to zero. The right panel shows a clock transition lineshape for unstabilized magnetic field with red circles and a much improved lineshape under stabilized field with blue filled circles. The gradual transition of the lineshape is depicted in the left panel via the transformation of the marker from a red circle to a filled blue circle. **c,** Measurements of DC electric field induced Stark shift on the clock transition show a clear quadratic behavior. The red circles show that a residual shift due to the stray DC

field can be as large as -1.3×10$^{-16}$. The blue squares show a much reduced shift after having purged the vacuum windows with N$_2$ gas. Dashed lines show a quadratic fit to the data. Solid vertical lines show the locations of a zero net electric field.

**Table 1**

**Shifts and Uncertainties in Fractional Frequency Units ×10$^{-18}$**

| Sources for Shift | $\Delta_{SrI}$ | $\sigma_{SrI}$ | $\Delta_{SrII}$ | $\sigma_{SrII}$ |
|---|---|---|---|---|
| BBR Static | -4832 | 45 | -4962.9 | 1.8 |
| BBR Dynamic | -332 | 6 | -346 | 3.7 |
| Density Shift | -84 | 12 | -4.7 | 0.6 |
| Lattice Stark | -279 | 11 | -461.5 | 3.7 |
| Probe Beam AC Stark | 2 | 5 | 0.8 | 1.3 |
| 1$^{st}$ Order Zeeman | 0 | <0.1 | -0.2 | 1.1 |
| 2$^{nd}$ Order Zeeman | -175 | 1 | -144.5 | 1.2 |
| Residual Lattice Vector Shift | 0 | <0.2 | 0 | <0.2 |
| Line Pulling & Tunneling | 0 | <0.1 | 0 | <0.1 |
| DC Stark | -4 | 4 | -3.5 | 2 |
| Background Gas Collisions | 0.07 | 0.07 | 0.63 | 0.63 |
| AOM Phase Chirp | -7 | 20 | -1 | 1 |
| 2$^{nd}$ Order Doppler | 0 | <0.1 | 0 | <0.1 |
| Servo Error | 1 | 4 | 0.4 | 0.6 |
| **Totals** | **-5710** | **53** | **-5922.5** | **6.4** |

**Table 1. Frequency Shifts and Related Uncertainties for SrI and SrII.** Errors are quoted as 1σ standard errors inflated by the square root of the reduced $\chi^2$ with statistical and systematic errors added in quadrature. The coefficient for the BBR dynamic correction was assumed to be a simple mean between the two most recent published values [28,38]. The error in this coefficient stands alone as the one non-statistical error in the SrII uncertainty budget. Background gas collisional shifts are estimated from the measurements reported in Ref. 39.

**Table 2**

**Probe Temperature Corrections and Uncertainties in mK**

| Corrections | $\Delta T$ | $\sigma_T$ |
|---|---|---|
| Calibration (including self-heating) | 0 | 16 |
| Residual Conduction | 0 | 0.7 |
| Temperature Gradient | 40 | 20 |
| Lead Resistance | 7.7 | 1.5 |
| Lattice Light Heating | -15 | 7.5 |
| **Totals** | 32.7 | 26.7 |

**Table 2. Systematic Uncertainty for the In-vacuum Silicon Diode Thermometer.** All errors are quoted as 1σ standard errors. The absolute calibration of the silicon diode sensor (including the self-heating effect) was performed by the vendor and the calibration is traceable to the NIST blackbody radiation standard. We have evaluated corrections and their uncertainties for the operation of the sensors in our vacuum chamber, including the residual conduction by the mount, extra lead resistance, and lattice light heating.

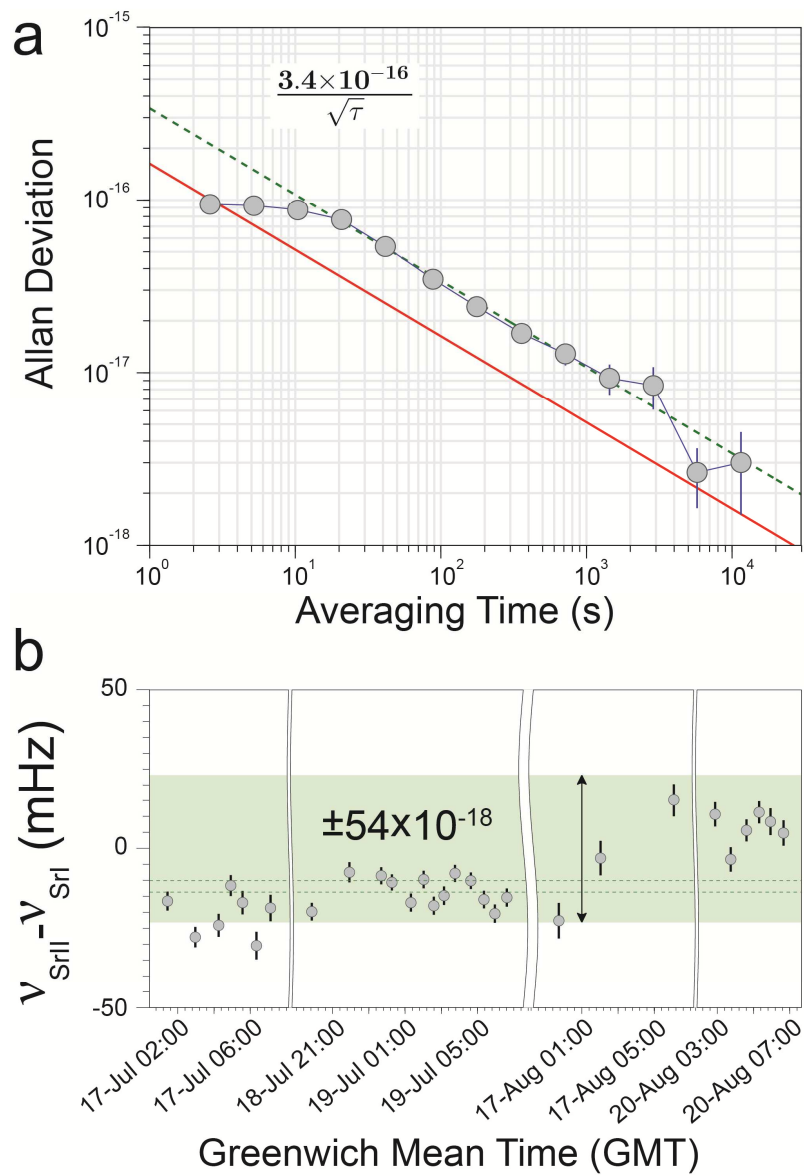

Fig. 1

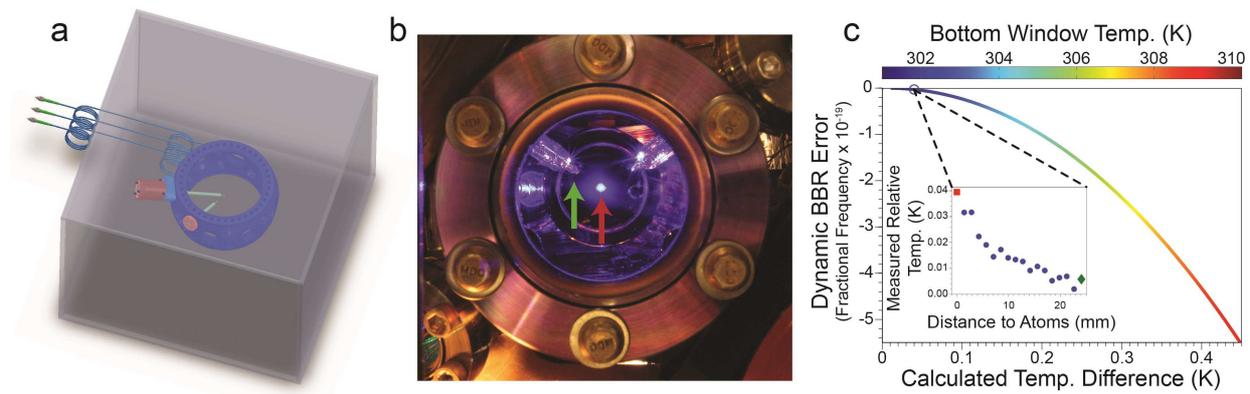

Fig. 2

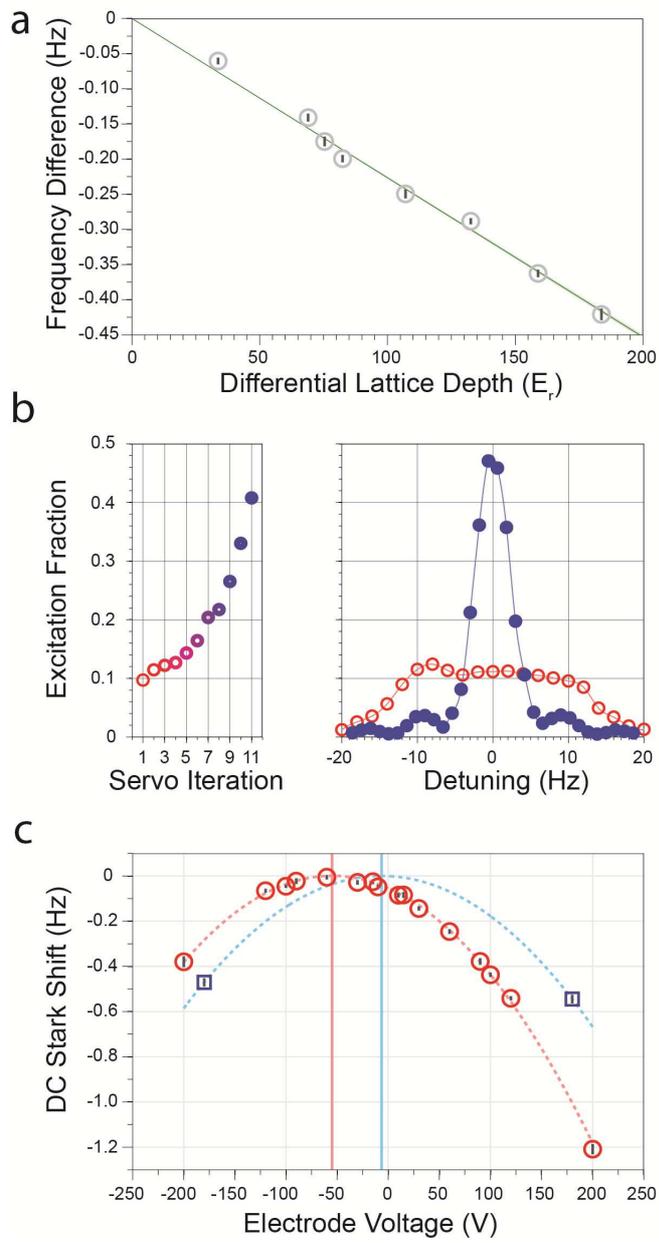

Fig. 3